\documentclass[aps,prb,reprint,amsmath,amssymb,superscriptaddress,longbibliography]{revtex4-1}
\usepackage{xcolor}
\usepackage{graphicx}
\usepackage{dcolumn}
\usepackage{bm}
\usepackage[normalem]{ ulem }
\usepackage{soul}
\usepackage{textcomp}

\usepackage{xcolor}
\usepackage{graphicx}
\usepackage{dcolumn}
\usepackage{bm}
\usepackage[normalem]{ ulem }
\usepackage{soul}
\usepackage[free-standing-units]{siunitx}

\newcommand{\nm}[1]{\SI{#1}{\nano\metre}}
\newcommand{\kHz}[1]{\SI{#1}{\kilo\hertz}}
\newcommand{\MHz}[1]{\SI{#1}{\mega\hertz}}

\definecolor{jl}{HTML}{FF0000}
\definecolor{mba}{HTML}{FF7F00}
\definecolor{av}{HTML}{4daf4a}

\begin{document}

\title{Cell selective manipulation with single beam acoustical tweezers.}

\author{Michael Baudoin}
\email{Corresponding author: \mbox{michael.baudoin@univ-lille.fr}}
\homepage{\mbox{http://films-lab.univ-lille1.fr/michael}}
\affiliation{Univ. Lille, CNRS, Centrale Lille, Yncréa ISEN, Univ. Polytechnique Hauts-de-France, UMR 8520 - IEMN, SATT NORD, F- 59000 Lille, France.}%
\affiliation{Institut Universitaire de France, 1 rue Descartes, 75005 Paris}%
\author{Jean-Louis Thomas}
\affiliation{Sorbonne Universit\'{e}, CNRS, Institut des NanoSciences de Paris, INSP, F-75005 Paris, France}%
\author{Roudy Al Sahely}%
\affiliation{Univ. Lille, CNRS, Centrale Lille, Yncréa ISEN, Univ. Polytechnique Hauts-de-France, UMR 8520 - IEMN, SATT NORD, F- 59000 Lille, France.}%
\author{Jean-Claude Gerbedoen}%
\affiliation{Univ. Lille, CNRS, Centrale Lille, Yncréa ISEN, Univ. Polytechnique Hauts-de-France, UMR 8520 - IEMN, SATT NORD, F- 59000 Lille, France.}%
\author{Zhixiong Gong}%
\affiliation{Univ. Lille, CNRS, Centrale Lille, Yncréa ISEN, Univ. Polytechnique Hauts-de-France, UMR 8520 - IEMN, SATT NORD, F- 59000 Lille, France.}%
\author{Aude Sivery}%
\affiliation{Univ. Lille, CNRS, Centrale Lille, Yncréa ISEN, Univ. Polytechnique Hauts-de-France, UMR 8520 - IEMN, SATT NORD, F- 59000 Lille, France.}%
\author{Olivier Bou Matar}%
\affiliation{Univ. Lille, CNRS, Centrale Lille, Yncréa ISEN, Univ. Polytechnique Hauts-de-France, UMR 8520 - IEMN, SATT NORD, F- 59000 Lille, France.}%
\author{Nikolay Smagin}%
\affiliation{Univ. Lille, CNRS, Centrale Lille, Yncréa ISEN, Univ. Polytechnique Hauts-de-France, UMR 8520 - IEMN, SATT NORD, F- 59000 Lille, France.}%
\author{Peter Favreau}%
\affiliation{Univ. Lille, CNRS, Centrale Lille, Yncréa ISEN, Univ. Polytechnique Hauts-de-France, UMR 8520 - IEMN, SATT NORD, F- 59000 Lille, France.}%
\author{Alexis Vlandas}
\email{Corresponding author: \mbox{alexis.vlandas@iemn.fr}}
\affiliation{Univ. Lille, CNRS, Centrale Lille, Yncréa ISEN, Univ. Polytechnique Hauts-de-France, UMR 8520 - IEMN, SATT NORD, F- 59000 Lille, France.}%

\date{\today}

\begin{abstract}
Acoustical tweezers open major prospects in microbiology for cells and microorganisms contactless manipulation, organization and mechanical properties testing since they are biocompatible, label-free and can exert forces several orders of magnitude larger than their optical counterpart at equivalent wave power  \cite{arfm_baudoin_2020}. Yet, these tremendous perspectives have so far been hindered by the absence of selectivity of existing acoustical tweezers \cite{tran2012,ding2012} - i.e., the ability to select and move objects individually - and/or their limited resolution restricting their use to large particle manipulation only  \cite{marzo2015,baresch2016,nature_melde_2016,prap_riaud_2017,sa_baudoin_2019,prap_lirette_2019}. Here, we report precise selective contactless manipulation and positioning of human cells in a standard microscopy environment, without altering their viability. Trapping forces of up to $\sim 200$ pN are reported with less than $2$ mW of driving power. The unprecedented selectivity, miniaturization and trapping force are achieved by combining holography with active materials and fabrication techniques derived from the semi-conductor industry to synthesize specific wavefields (called focused acoustical vortices \cite{hefner1999,arfm_baudoin_2020}) designed to produce stiff localized traps. We anticipate this work to be a starting point toward widespread applications of acoustical tweezers in fields as diverse as tissue engineering \cite{Eifeng2018}, cell mechano-transduction analysis \cite{block1989compliance,Suresh2007,Ming2010}, neural network study \cite{AEBERSOLD201660} or mobile microorganisms imaging \cite{Thalhammer2011,Ahmed2016}, for which precise manipulation and/or controlled application of stresses is mandatory.

\end{abstract}

\maketitle

\section{\label{sec:introduction}Introduction}
 Contactless tweezers based on optical \cite{ashkin1986,grier2003revolution,Neuman2004} and magnetic forces \cite{ecr_crick_1950,s_smith_1992,s_strick_1996} have been developed in the last decades and have led to tremendous progress in science recognized by several Nobel prizes. Nevertheless, these technologies have stringent limitations when operating on biological matter. Optical tweezers rely on the optical radiation pressure, a force proportional to the intensity of the wavefield divided by the speed of light. The high value of the latter severely limits the forces that can be applied and imposes the use of high intensity fields. This can lead to deleterious photothermal damages (due to absorption induced heating) and/or photochemical damages (due to excitation of reactive compounds like singlet oxygen)\cite{Neuman1999,bj_liu_1995,bj_liu_1996,Blazquez2019} adversely affecting cells integrity. Magnetic tweezers, on the other hand, can only manipulate objects susceptible to magnetic fields and thus require other particles to be pre-tagged with magnetic compounds, a limiting factor for many applications. For biological applications, acoustical tweezers are a prominent technology  \cite{csr_lenshof_2010,nm_sitters_2015,nc_collins_2015,nm_ozcelik_2018,arfm_baudoin_2020}. They rely on the acoustical radiation force \cite{loc_bruus_2012,arfm_baudoin_2020}, which is -as for their optical counterpart- proportional to the intensity of the wave divided by the wave speed. But, the dramatically lower speed of sound compared to light leads to driving power several orders of magnitude smaller than in optics to apply the same forces (or conversely, forces several orders of magnitude larger at the same driving power) \cite{baresch2016,Zhao2019}. In addition, the innocuity of ultrasounds on cells and tissues below cavitation threshold is largely documented \cite{umb_stuart_2000,hultstrom2007,wiklund2012,po_burguillos_2013,marx2015} and demonstrated daily by their widespread use in medical imaging \cite{ap_szabo_2014}. Indeed, the frequencies typically used in ultrasound applications (\kHz{100} to \MHz{100}) are far below electronic or molecular excitation resonances thus avoiding adverse effects on cells integrity. Moreover, the weak attenuation of sound in both water and tissues at these frequencies limits absorption induced thermal heating. Finally, almost any type of particles (solid particles, biological tissues, drops) can be trapped without pre-tagging \cite{nm_ozcelik_2018} and the low speed of sound enables spatial resolution down to micrometric scales even at these comparatively low frequencies.

\begin{center}
\begin{figure*}[htbp]
\includegraphics[width=\textwidth]{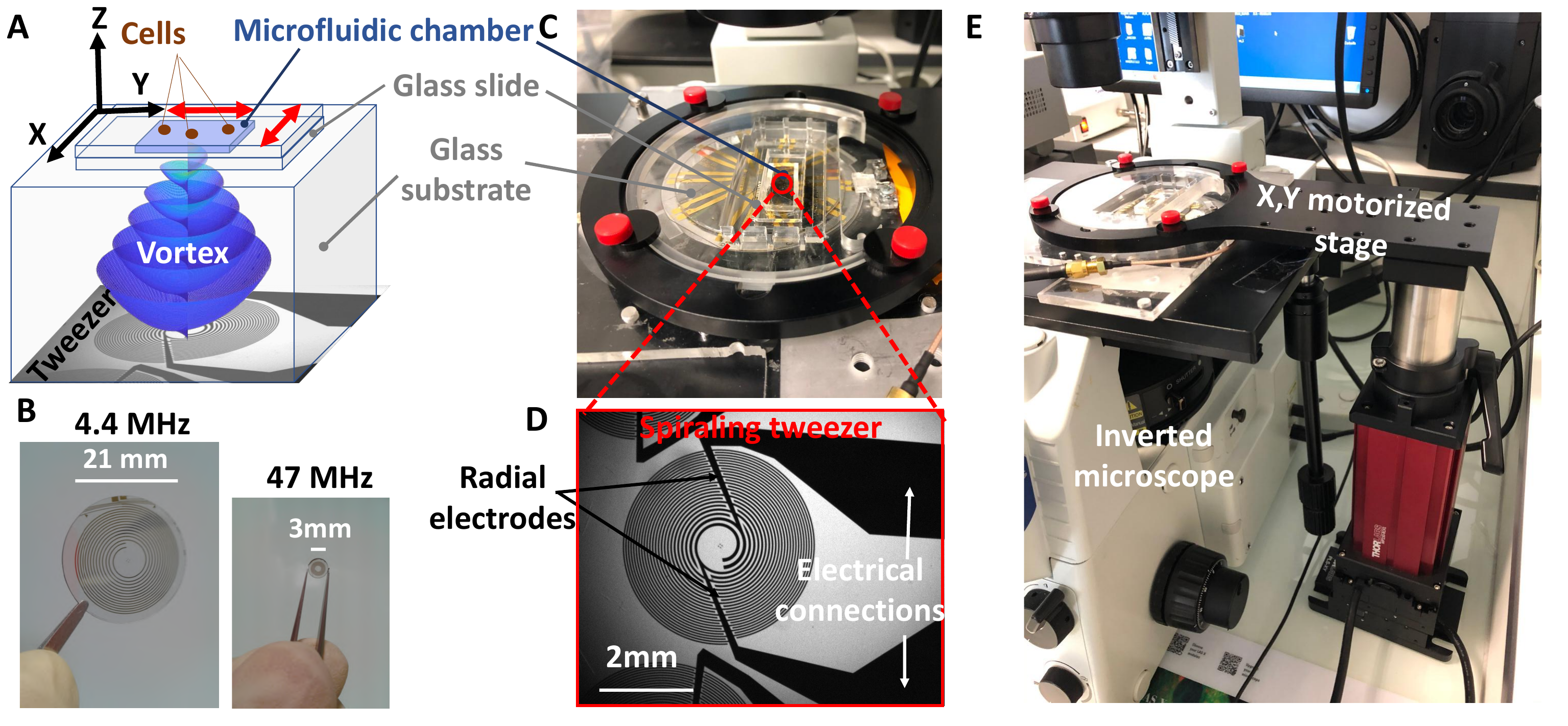}
\caption{Experimental setup. A. Illustration of the working principle of the tweezers designed for cells selective manipulation: A spherically focused acoustical vortex is synthesized by spiraling active electrodes metallized at the surface of a piezoelectric substrate and actuated with a function generator connected to an amplifier. The vortex propagates and focalizes inside a glued glass substrate and then reaches a microfluidic chamber made of a glass slide and a PDMS cover containing cells embedded in a growth medium. The microfluidic device is acoustically coupled with the transducer with a thin layer of silicone oil (25 cSt). A cell located at the center of the acoustical vortex is trapped. Its motion relative to other cells is enabled by the displacement of the microfluidic chamber driven by a XY motorized stage (See Movie 1 for an animated explanation of the setup working principle). B. Picture of typical transducers used in the present study (right) and illustration of the scale reduction compared to previous lower frequency designs by Baudoin \textit{\textit{et al}}. \cite{sa_baudoin_2019} (left). C. Image of the actual experimental setup. D. Zoom in on the spiral transducer and the electrical connections (in black).  E. Illustration of the integration of the whole setup inside a standard inverted microscope. Photo credit: B: J.-C. Gerbedoen, SATT NORD / C-D-E: R.A. Sahely, Univ. Lille.}
\label{figure1}
\end{figure*}
\end{center}

Nevertheless, the promising capabilities offered by acoustical tweezers have so far been hindered by the lack of selectivity of existing devices and/or their restricted operating frequency limiting their use to large particles only. Yet, the ability to select, move and organize individual microscopic living organisms is of the utmost importance in microbiology for fields at the forefront of current research such as single cell analysis, cell-cell interaction study, or to promote the emergence of disruptive research e.g. on spatially organized co-cultures. In this paper, we unleash the potential of acoustical tweezers by demonstrating individual biological cells manipulation and organization in a standard microscopy environment with miniaturized single beam acoustical tweezers. The strength and efficiency of acoustical tweezers is illustrated by exerting forces on cells one order of magnitude larger than the maximum forces reported with optical tweezers \cite{m_keloth_2018} ($\sim 200$ pN), obtained with one order of magnitude less wave power ($<2$ mW).

\section{\label{sec:atd} Acoustical tweezers design}

The first experimental evidence of large particles trapping with acoustic waves dates back to the early 20th century \cite{sp_boyle_1928}. Nevertheless, the first demonstration of controlled manipulation of micrometric particles and cells with acoustic waves appeared only one century later with the emergence of microfluidics and high frequency transducers based on interdigitated electrodes \cite{tran2012,ding2012}. In these recent works, trapping relies on the 2D superposition of orthogonal plane standing waves, an efficient solution for the collective motion of particles, but one which precludes any selectivity, i.e., the ability to select and move one particle out of a population \cite{arfm_baudoin_2020}. Indeed, the multiplicity of nodes and anti-nodes leads to the existence of multiple trapping sites \cite{prap_silva_2019} which cannot be moved independently. In addition, multiple transducers or reflectors positioned around the manipulation area are mandatory for the synthesis of standing waves, a condition difficult to fulfill in many experimental configurations.

Selective trapping with single beam requires strong spatial localization and hence tight focusing of the wavefield. In optics, this ability has been achieved with focused progressive waves \cite{ashkin1986}, a solution also investigated in acoustics \cite{lee2009}. But such wavefields are inadequate in acoustics for most particles of practical interest, since objects with positive contrast factors (such as rigid particles or cells) are attracted to pressure nodes \cite{gorkov1962,loc_bruus_2012} and would be expelled from the focal point of a focused wave \cite{jasa_baresch_2013}. Acoustical vortices \cite{hefner1999}  provide an elegant solution to this problem \cite{jap_barech_2013}. These focused helical progressive waves spin around a central axis wherein the pressure amplitude vanishes, surrounded by a ring of high pressure intensity, which pushes particles toward the central node. Two-dimensional trapping \cite{courtney2014,prap_riaud_2017} and three-dimensional levitation  \cite{marzo2015} and trapping \cite{baresch2016} have been previously reported at the center of laterally and spherically focused vortices, respectively. However, all these demonstrations were performed on relatively large particles ($> $300 $\mu$m in diameter) using complex arrays of transducers, which are cumbersome, not compatible with standard microscopes, and that cannot be easily miniaturized to trap micrometric particles. Recently, Baudoin \textit{et al}. \cite{sa_baudoin_2019} 
demonstrated the selective manipulation of 150 $\mu$m particles in a standard microscopy environment with flat, easily integrable, miniaturized tweezers. To reach this goal, they sputtered holographic electrodes at the surface of an active piezoelectric substrate, designed to synthesize a spherically focused acoustical vortex. 

Nevertheless, transcending the limits of this technology to achieve selective cells manipulation remained a major scientific and technological challenge. Indeed, the system should be scaled down (frequency up-scaling) by a factor of 10 (since cells have typical size of 10 $\mu$m), while increasing dramatically the field intensity, owing to the low acoustic contrast (density, compressibility) between cells and the surrounding liquid \cite{pre_settnes_2012,natcom_augustsson_2016}. In addition, since the concomitant system's miniaturization and power increase are known to adversely increase the sources of dissipation, the tweezers had to be specifically designed to prevent detrimental temperature increase and enable damage free manipulation of cells: 

First, spherically focused acoustical vortices (Fig. \ref{figure1}A) were chosen to trap the particles. Indeed, the energy concentration resulting from the 3D focalization (Fig. \ref{figure2}F) enables to reach high amplitudes at the focus from remote low power transducers. These spherically focused vortices were synthesized by materializing the  hologram of a  \MHz{\sim 45} vortex   \cite{sa_baudoin_2019} with metallic electrodes at the surface of an active piezoelectric substrat. The hologram was discretized on two levels resulting in two intertwined spiralling electrodes (Fig. \ref{figure1}D), patterned in a clean room by standard photo-lithography techniques (see Methods section A). 
The scale reduction compared to our previous generation of acoustical tweezers \cite{sa_baudoin_2019} is illustrated in Fig. \ref{figure1}B. Second, the design of the electrodes was optimized to reduce Joule heating (magnified by the scale reduction) inside the electrodes. To prevent this effect, (i) the thickness of the metallic electrodes was increased by a factor of 2 (\nm{400} of gold and \nm{40} of titanium); (ii) the width of the electrical connections (Fig. \ref{figure1}D) supplying the power to the spirals was significantly increased to prevent any dissipation before the active region; and (iii) two radial electrodes spanning half of the spirals were added as a way to effectively bring power to the driving electrode. Third, a $1.1$ mm glass substrate (Fig. \ref{figure1}A, \ref{figure1}C) was glued to the electrodes and placed in between the transducers and the microfluidic chamber wherein the cells are manipulated. This glass substrate has a double function: (i) it enables the focalization of the wave and (ii) it thermally insulates the microfluidic device from the electrodes thanks to the poor thermal conductivity of glass.

The final device hence consists of (see Movie 1 in SI, Fig. \ref{figure1}A, \ref{figure1}C, \ref{figure1}E): (i) spiralling holographic transducers generating an acoustical vortex which propagates and focuses inside a glass substrate ; (ii) a microfluidic PDMS chamber supported by a glass slide containing cells and placed on top of the substrate, wherein the acoustical vortex creates a trap and (iii) a motorized stage that enables the X,Y displacement of the microfluidic chamber with respect to the trap. The whole transparent setup is integrated in an inverted microscope as depicted in Fig. \ref{figure1}E. 

\section{\label{sec:cat} Characterization of the acoustical trap}

\begin{center}
\begin{figure*}[htbp]
\includegraphics[width=\textwidth]{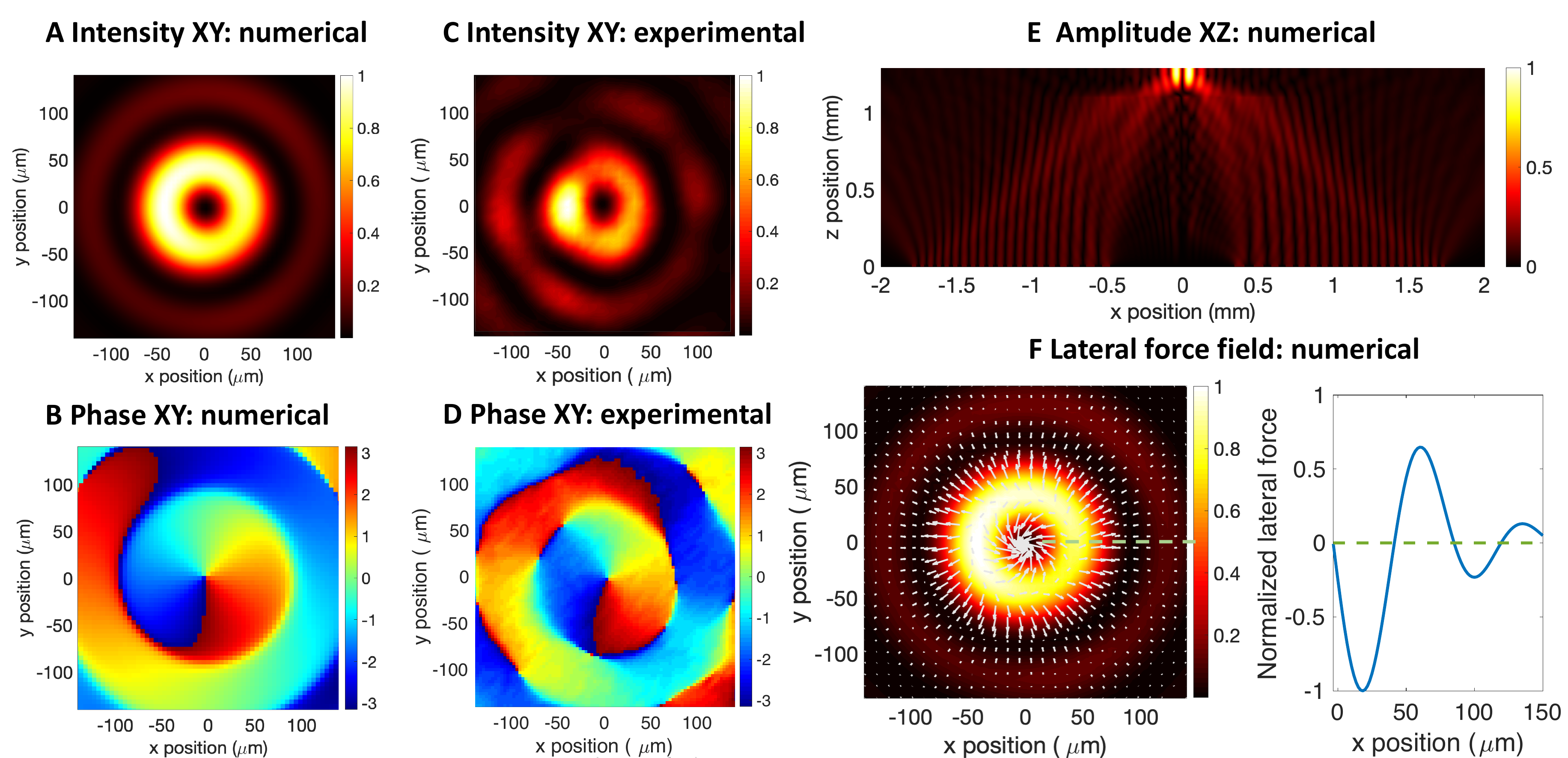}
\caption{Acoustic field and radiation forces. A-D. Numerical predictions (A and B) and experimental measurements (C and D) with a UHF-120 Polytec laser Doppler vibrometer of the normalized modulus (A and C) and phase (C and D) of the acoustic normal displacement at the surface of the glass slide (XY plane). In the experiments presented in this paper, the maximum amplitude of the vibrations (displacement) on the ring lies typically between $0.1$ nm and $1$ nm depending on the electrical power applied to the transducers. E. Simulated evolution of the amplitude of the acoustic field in the propagation plane (XZ). This simulation illustrates the concentration of the acoustic energy through focalization. F. Normalized magnitude and distribution of acoustic forces. Left: the white arrows show the convergence of the force field toward the center of the beam but also that the first ring is repulsive for particles located outside the trap. Right: Magnitude of the lateral force along the green dashed line plotted on the left figure. When the force is negative, the particle is pushed toward the center of the acoustic vortex, while when it is positive it is pushed outward. Zero values correspond to static equilibrium positions. The magnitude of the maximum trapping force computed with the code varies between $30 \, $pN and $650 \, $pN (see Methods section D) for vibration amplitude of $1$ nm (acoustic power of 2 mW) depending on the exact cells acoustic properties \cite{pre_settnes_2012,natcom_augustsson_2016}}.
\label{figure2}
\end{figure*}
\end{center}

The principle of high frequency acoustical vortices synthesis with these active holograms was assessed through the comparison of numerical predictions obtained from an angular spectrum code and experimental measurements of the acoustic field normal displacement at the surface of the glass slide (XY plane) with a Polytech UHF-120 laser Doppler vibrometer (Fig. \ref{figure2}, A-D). Both the intensity and phase are faithful to the simulations and demonstrate the ability to generate high frequency acoustic vortices. As expected, the wavefield exhibits a central node (corresponding to the phase central singularity) surrounded by a ring of high intensity which constitutes the acoustical trap. The magnitude of the acoustic field (displacement) depends on the driving electrical power and was measured to vary typically between $0.1$ nm and $1$ nm, at the electrical power used in the manipulation experiments. This corresponds to acoustic powers lying between $20 \, \mu $W and $2$ mW  (see Methods section I). The concentration of the acoustic energy through focalization in the propagation plane (XZ) can be seen in Fig. \ref{figure2}E. 

An estimation of the lateral force field exerted on a cell of 10 $ \mu $m radius with density 1100 kg m$^{-3}$ and compressibility $4\times10^{-10}$ Pa$^{-1}$ was computed at each point in the manipulation plane of the microfluidic chamber (XY plane, Fig. \ref{figure2}F) with the theoretical formula derived by Sapozhnikov \& Bailey \cite{jasa_sapozhnikov_2013}.
This calculation gives an estimation of the force of the order of $100 $ pN, which can nevertheless strongly vary depending on the exact cells properties (see Methods section D for the exact values depending on cells acoustic properties\cite{pre_settnes_2012,natcom_augustsson_2016} for an acoustic vibration of $1 \, $nm).

Finally, the temperature increase due to Joule heating in the electrodes as well as the total temperature increase due to both Joule heating and acoustic wave absorption was measured using  an infrared camera to assess potential impact on biological material (See Methods section J). For most experiments presented in this paper (corresponding to acoustic displacement $< 0.6$ nm), the temperature increase is lower than $2.2^{\circ}$C after 2 min of manipulation and even vanishes for the lowest power ($0.1 \,$nm). It reaches a maximum value of $5.4^{\circ}$C at the top of the glass slide and $5.5^{\circ}$C inside a drop of glycerol placed on top of the glass slide (acting as a perfectly absorbing medium) at the highest power used for high speed  displacement of the cells. These measurements indicate that the first source of heat is Joule heating in the electrodes which could be solved by active cooling of the transducer. They also suggest that even at the largest power used in the present experiments, the moderate temperature increase remains compatible with cells manipulation, as assessed in the next section.

\section{\label{sec:cmp} Cells manipulation, positioning and viability}

\begin{figure*}[htbp]
\includegraphics[width=\textwidth]{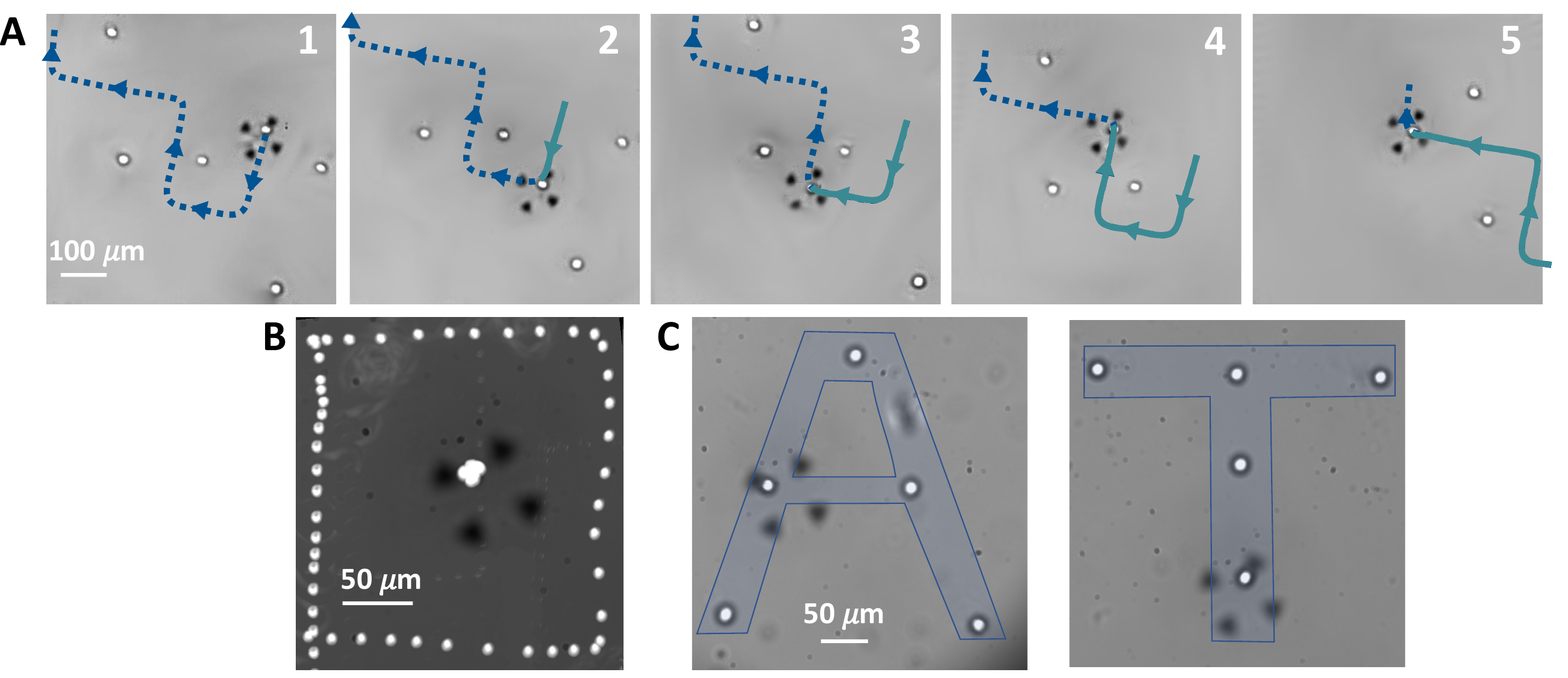}
\caption{A. Stack of images illustrating the selective manipulation of a human breast cancer cell (MDA-MB-231) of radius 7 $\pm  1 $ $\mu$m between other cells. The blue dotted line and green continuous line show respectively the future and past path followed by the cell. (See also Movie 2) B. Image illustrating the square relative motion of a trapped cell "1" of 7 $\pm  1 $ $\mu$m (located in the center of the picture) around another cell "2" obtained by superimposing the images of the two cells in the frame of reference of the trapped cell (see also Movie 3). In this frame of reference, the successive positions of cell "2" form a square. For the sake of clarity other cells appearing in the field of view have been removed. C. Manipulation of 10 MDA cells (average radius $9 \, \mu $m to form the letters "A" and "T" of "Acoustical Tweezers". Note that in these pictures the focus is voluntarily left under-focused to improve contrast of the cells.}   
\label{figure3}
\end{figure*}

Cell manipulation is demonstrated in a microfluidic device integrated in a standard inverted microscope (Fig. \ref{figure1}E) to illustrate the fact that our approach can be easily transposed to standard microbiology experiments. The device is composed of a thin glass slide treated to prevent cell adhesion and a PDMS chamber of controlled height (38 $\mu$m). The cells are loaded by placing a drop  of the cell suspension (10-20 $\mu$L) on the glass surface using a micro-pipette and carefully lowering the chamber on top of the drop. The position of the vortex core is spotted with four triangular marks deposited at the surface of the glass substrate. Using a XY positioning system it is thereafter possible to align the tweezers center to any cell present in the chamber. Upon activation of the AC driving signal, a cell situated inside the vortex core is nearly instantaneously trapped.

\begin{figure*}[htbp]
\includegraphics[width=0.8\textwidth]{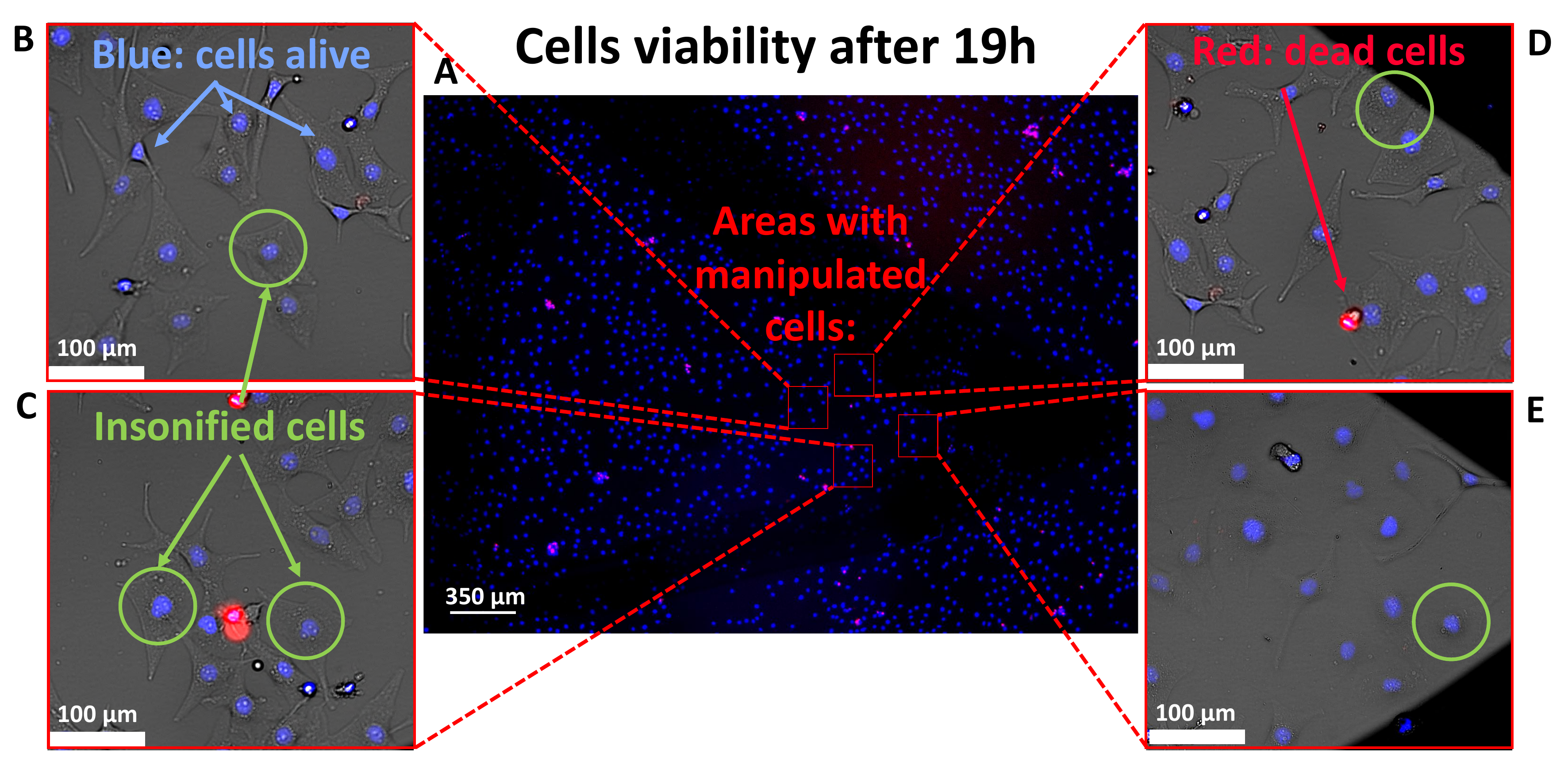}
\caption{A. Overview of the central part of the microfluidic device in which the viability experiments were performed. The cells are stained using a viability kit and imaged at $360$ nm and $535$ nm excitation ($460 $ nm and $617 $ nm emission). The cell nucleus are represented in blue, while the dead cells appear in red. The whole field of view contains 4581 cells (226 dead - 5\%) while the region where manipulation took place contains 166 cells  (5 dead - 3\%).
B-E Details of the 5 cells exposed to the acoustical tweezers for 2 min (4 others were exposed on another similar device). The green circle represents the first ring of the trap.}
\label{figure4}
\end{figure*}

The first demonstration of the selective nature of our tweezers is showcased by our ability to pick up a single cell (breast cancer cell MDA-MB-231, $7 \pm 1 \, \mu m$ in radius) amongst a collection of cells and move it along a slalom course where other free to move cells act as poles (see Fig. \ref{figure3}A and Movie 2). Then a second cell initially serving as a slalom marker, is moved to prove that it was free (Movie 2). The precise displacement can be performed in any direction as demonstrated by the square motion of a cell around another (Fig. \ref{figure3}B, Movie 3).  
Displacement can be performed even in the presence of other cells without any risk of "coalescence" as the first ring acts as a barrier. As can be seen in Fig. \ref{figure2}C, the radius of the first repulsive ring is typically 40 $\mu$m. The second ring of much weaker intensity can also slightly affect free cells at large power.

One of the key ability enabled by acoustical tweezers is the capture, positioning and release of cells at precise locations. As an illustration, a total of 10 individual MDA cells were therefore positioned to spell the letter "A" and "T" of "Acoustical Tweezers" (Fig. \ref{figure3}C). The total manipulation time to achieve these results was kept under 10 min (less than 2 min per cell). All the operations represented in Fig. \ref{figure3} were performed with acoustic vibration displacements $<0.5 \, $nm. 

Finally, we performed some experiments to quantify the forces that can be exerted on cells with these tweezers. For this purpose, cells were trapped and then moved with an increasing speed until it was ejected from the trap. Velocities up to 1.2 mm$\,$s$^{-1}$ before ejection have been measured for cells displacement of diameter $12 \pm1 \, \mu$m trapped with an acoustic field of magnitude $0.9 \,$nm  in a micro-chamber of height $38 \, \mu$m (see Movie 4). This corresponds to a trapping force of $194 \pm 35$ pN according to Faxen's formula \cite{s_happel_1965} (see Method section E), which lies in the range predicted by theory in section \ref{sec:cat}.  As a comparison, this force is one order of magnitude larger than the maximum forces (20 pN) reported by Keloth \textit{et al}. \cite{m_keloth_2018} with optical tweezers and obtained with one order of magnitude less power ($1.8$ mW here compared to the $26.8$ mW used for optical trapping. Furthermore, unlike with optical tweezers, it is still possible to substentially increase this force with acoustical tweezers by increasing the actuation power and improving the thermal management of the device, as most of the dissipated power comes from the transducer and not from the direct absorption by the medium.

As described in the introduction, one of the main gains which can be expected from transitioning from optical to acoustical tweezers is the absence of deleterious effects of the latter when manipulating living cells. The short-term and long-term viability was investigated using a fluorescent viability assay as well as post exposure cell observation. A first set of experiments was thus conducted to address the short-term viability of MDA cells. The cells were captured for 2 min in the vortex at maximum power (amplitude 0.9 nm) to mimic a standard positioning sequence and observed for any sign of damage during manipulation and for 30 min afterwards. During manipulation, no increase of fluorescence was observed suggesting that the sound field does not induce membrane permeabilization which often correlates with viability decrease \cite{Rols2017}. After the tweezers were switched off, the cell did not display any increase of fluorescence and remained at an intensity well under the dead cells found nearby (5$\times$ to 10$\times$ lower, see SI). This strongly supports that short-term damages produced by the acoustical tweezers is minimal. 

It is however known that damages experienced by a cell can lead to its death for hours afterwards \cite{ref_damage_laser}. To assess the long-term impact of cell manipulation using acoustical tweezers, we performed a viability assay overnight. The MDA cells were seeded at 60 (\%) confluence ratio in two glass devices with no surface treatment and left to re-adhere for 5h. Nine cells located at different positions in the two different microfluidic chambers were exposed to the tweezers of acoustic vortex at maximum power for 2 min each. An observation of the cells was performed after 19h (half the population doubling rate of MDA cells \cite{ATCC}) to compare their viability with a control region of the device (see Fig. \ref{figure4}A).  No extra mortality was observed in the illuminated region (dead/live cell ratio of 3\%) compared to the statistics performed on the overall device (dead/live cell ratio of 5\%). This likely indicates that the dead cells are depositing randomly and that the tweezers do not provoke extra mortality. We also studied in detail the fate of the nine illuminated individual cells (see Fig. \ref{figure4}B-E). All the cells exposed to the acoustic field (the green circle indicates the extension of the first ring of the vortex) and their immediate neighbours were alive and showed no difference compared to the nearby cells.  

\section{\label{sec:d} Conclusion and outlooks}

In this work, cell harmless selective manipulation is demonstrated through the capture and precise positioning of individual cells amongst a collection in a standard microscopy environment. Both short-term and long-term viability of manipulated cells is evaluated, showing no impact on cells integrity. This opens widespread perspectives for biological applications wherein precise organization of cells or microorganisms is a requisite. In addition, trapping force over wave intensity ratio two orders of magnitude larger than the one obtained with optical tweezers is reported with no deleterious effect such as phototoxicity. Further engineering optimization of these tweezers to limit Joule heating will hence enable the application of stresses several orders of magnitude larger than with optical tweezers without altering cells viability, a promising path for acoustic spectroscopy \cite{nm_sitters_2015}, cell adhesion \cite{nrmcb_parsons_2010} or cell mechano-transduction \cite{block1989compliance,Suresh2007,Ming2010} investigation. In addition, new abilities could be progressively added to these tweezers: The focused vortex structure used for selective particle trapping in this paper is also known to exhibit 3D trapping capabilities \cite{jap_barech_2013,baresch2016}. This function was not investigated here owing to the confined nature of the microchamber but could closely follow this work. Synchronized vortices could also be used to assemble multiple particles, as recently suggested by Gong \& Baudoin \cite{prap_gong_2019}. This would enable the investigation of tissue engineering and envision 3D cell printing. Finally, the most thrilling and challenging perspective to this work might be the future development of Spatial Ultrasound Modulators (analogs to Spatial Light Modulator in optics), designed to manipulate and assemble many objects simultaneously. While such a revolution is on the way for large particles manipulation in air \cite{prl_poulikakos_2014,pnas_marzo_2019,n_hirayama_2019}, it would constitute a major breakthrough at the microscopic scale in liquids wherein the actuation frequencies are 3 orders of magnitude larger. The present work hence constitutes the cornerstone towards widespread applications of acoustical tweezers for biological applications.

\end{document}